\documentclass[twocolumn,showpacs,preprintnumbers,amsmath,amssymb,prl]{revtex4}
\usepackage{amsmath}
\usepackage{amssymb}
\usepackage{hyperref}
\usepackage{graphicx}

\newcommand\gcl{{g_{\rm cl}}}
\newcommand\gcand{{g_{\rm evap}}}
\newcommand\editremark[1]{{ }}
\newcommand\optional[1]{}

\newcommand\unitkms{ \,{\rm km}\,{\rm s}^{-1}}

\newcommand\SN{SN{} }   % abbreviation
\newcommand\GW{GW{} }   % abbreviation

\begin{document}

\title{Dynamical Interactions and the Black Hole Merger Rate of the Universe}% Force line breaks with \\

\author{Ryan M. O'Leary}
\email{roleary@cfa.harvard.edu}
\affiliation{%
Harvard-Smithsonian Center for Astrophysics, 60 Garden St., Cambridge,
MA 02138
}
\author{Richard O'Shaughnessy}%
\email{oshaughn@northwestern.edu}
\author{Frederic A. Rasio}
\email{rasio@northwestern.edu}
\affiliation{%
Northwestern University, Department of Physics and Astronomy, 2132 Tech Drive, Evanston, IL 60208
}%

\date{August 16, 2007}% It is always \today, today,

\begin{abstract}
Binary black holes can form efficiently in dense young stellar
clusters, such as the progenitors of globular clusters, via a combination of
gravitational segregation and cluster evaporation.  We use simple
analytic arguments supported by detailed $N$-body simulations to
determine how frequently  black holes born in a single stellar cluster
should form binaries, be ejected from the cluster, and merge through
the emission of gravitational radiation.  We then convolve this
``transfer function'' relating cluster formation to black hole mergers
with (i) the distribution of observed cluster masses
 and (ii) the star formation history of the universe,
assuming that a significant fraction $\gcl$ of star formation occurs in
clusters and that a significant fraction $\gcand$ of clusters undergo this
segregation and evaporation process.
 We predict future ground--based gravitational wave (GW) detectors could
observe $\sim 500 (\gcl/0.5) (\gcand/0.1)$  double black hole
mergers per year, and the presently operating LIGO interferometer
would have a chance (50\%) at detecting a merger during its first full
year of science data.    More realistically, 
advanced LIGO and similar next-generation gravitational wave
observatories provide unique opportunities to  constrain otherwise
inaccessible properties of  clusters formed in the early universe. 
\end{abstract}

%04.30.Db  Wave generation and sources
%95.55.Ym  Gravitational radiation detectors; mass spectrometers; and other instrumentation and techniques (see also 04.80.-y Experimental studies of gravity in general relativity and gravitation)
%97.60.Lf  Black holes (see also 04.70.-s Physics of black holes in general relativity and gravitation; for galactic black holes, see 98.35.Jk and 98.62.Js)
\maketitle

% POINT: Background/outline (required for letter style)
Given our understanding of how isolated binary stars evolve,
noninteracting stellar systems should produce relatively few double
black hole (BH-BH) binaries tight enough to merge through the emission of
\GW within the age of the universe 
\cite{2006astro.ph.12032B,PSconstraints,StarTrack}.
\citet{PZMcM} demonstrated that 
\emph{interactions} between  black holes (BHs) in dense
cluster environments could  produce  merging
BH-BH binaries much more efficiently than through the evolution of
isolated binaries.  As a result, the local binary black hole merger
rate -- the net rate both from isolated evolution of noninteracting
stars and from dense clusters -- can depend sensitively on the formation and
evolution of young clusters through the entire history of the universe.

% POINT: Stellar clusters significant fraction, possibly ubiquitous
An increasing number of 
galactic \cite{1991ApJ...371..171L,2003ARAA..41...57L} and 
extragalactic \cite{1998ASPC..148..127H,2005ApJ...619..779C}
observations suggest at least $20\%$ of stars form in dense, interacting
stellar clusters.
Over time, each cluster  dissipates, both because hot young stars and
supernovae (SN) heat and 
eject a significant fraction of the residual gas that gravitationally
binds the cluster (``infant mortality''), and because the host galaxy's tidal field  strips
off stars as the cluster orbits it
\cite{2001ApJ...561..751F,2001ApJ...550..691J}.  Thus any set of coeval   clusters
decrease  in number and size, spewing stars into their hosts 
\cite{2006ApJ...650L.111C,2005ApJ...631L.133F}, with only a few of the
most initially dense and orbitally-fortunate clusters surviving to the
present.  
\editremark{[For this reason the \emph{present-day} fraction of stars
in the Milky Way found in globular clusters is much less than unity.]}

% POINT: Unknown fraction...but can constrain with this paper!
Unfortunately, electromagnetic observations of large young clusters in
other galaxies cannot resolve their internal  structure.  These
observations therefore 
only weakly constrain  the  fraction of young clusters that
survive their first few Myr, during which the most massive young stars evolve,
supernovae, and give birth to black holes.
\editremark{FIND A BEST ESTIMATE for survival distribution and
  correlation with parameters - see
  the recent Fall paper?}
% POINT: Alternative channel
Similarly,  they cannot determine the efficacy of \emph{runaway stellar
collisions}, in which massive stars in particularly dense clusters
gravitationally segregate and collide to form very massive stellar
progenitors to large [$\gtrsim 100\, M_\odot$] single or binary black holes
\cite{1999AA...348..117P,PZ-Runaway,%
Ato-Runaway1,Ato-Runaway2,imbhlisa-2006}. \editremark{Provide simple
expression: ratio of core collapse time to nuclear evolution time -
want SN to be SLOW compared to collapse for runaway}
However, clusters give birth to many binary black holes, whose
\GW merger signals  provide unambiguous information
about the processes which produced them.  For example,
\citet{imbhlisa-2006} demonstrated that runaway stellar collisions
produce very massive [$\sim 100\, M_\odot$] double BH binaries often
enough to be easily and unambiguously  seen  with advanced ground-based interferometers
(i.e., advanced LIGO/VIRGO).   The detection rate of such massive BH
mergers  would therefore constrain the fraction of young clusters
which undergo collisional runaway.
% POINT : Here, GW for smaller clusters
Similarly, in this Letter we argue that a high \GW detection
rate of $\sim 15-20\, M_\odot$ black holes will unambiguously measure how often
young clusters  dynamically evaporate their most massive compact
components.  Specifically, we use the results of a recent
set of numerical simulations of BHs in clusters \cite{2006ApJ...637..937O} to
present a simple, analytic approximation relating the BH-BH
merger rate to properties of young clusters.
Existing observations only weakly constrain the first few Myr in the
lives of the most massive and ancient clusters; advanced ground-based
gravitational 
wave detectors therefore can 
explore new facets about the evolution of these
first massive clusters.

Within each
cluster, stars form according to an approximately power-law mass distribution
\citep{2005AA...430..941P}, with only a small number $N_{bh}\approx
3\times 10^{-3} (M_{cl}/M_\odot)$ of 
stars  more massive than $20 M_\odot$, roughly the mass needed to
guarantee a black hole forms in a \SN.
These most massive stars should rapidly
evolve, undergo \SN, and form black holes on a timescale much shorter
than the relaxation time of the cluster.
The most massive clusters with
$M_{cl}>M_{\rm crit}\equiv 3\times 10^4$ should contain more than $100$
such stars, each of which produces a roughly $m_{bh}\approx 10-25 M_\odot$ hole \cite{2004ApJ...611.1068B}. 
Being the most massive objects in the cluster, systems (single or binary) containing
BHs will rapidly mass segregate to the cluster core
\cite{2002ApJ...570..171F,Ato-Runaway1}.   In the process
exchange interactions will quickly break up any remaining star-BH
binaries \cite{ADM:Heg96}. \editremark{A better link on exchanges
  would be nice - this is just cross-sections.} 
Since the numerous black holes
significantly outweigh the average constituents of the cluster ($\left<m_{*}\right>\approx 0.5 M_\odot$), the
black holes can  decouple from the  stars in a process
known as the ``Spitzer instability''
\citep{1969ApJ...158L.139S,2000ApJ...539..331W}.  The more massive BHs 
interact and evolve on a more rapid timescale $\approx t_{cl}
\left<m_*\right>/\left<m_{bh}\right>$, quickly evaporating and
ejecting single and binary BHs.  Thermal equilibrium with the surrounding
stellar cluser is only restored  when so
few BHs remain that the subcluster's internal timescale once again
becomes commensurate with the cluster's interaction timescale.
This process of segregation, decoupling, and evaporation has been
extensively studied theoretically \cite{1993Natur.364..421K,1993Natur.364..423S} and
numerically, both in 
full $N$-body simulations 
($N\approx 10^3-10^5$) 
% Merrit et al (core formation by massive remnants)
\cite{2004ApJ...608L..25M,PZMcM,2006astro.ph..7461P,Spurzem-NbodySpitzerInstability} and in using
approximations in larger $N$ systems \cite{Ato-Runaway1,2006ApJ...637..937O}.

Previous studies have suggested that interactions in this dense
cluster can
produce many BH-BH mergers, whether through  evaporated binaries
\cite{PZMcM} 
or  through runaway BH-BH mergers in the dense cluster itself
\cite{2006ApJ...640..156G,MillerHamilton-BHCollisionRunaway2002,2002ApJ...566L..17M}.
However, these studies have faced significant objections to their
details, because each has omitted some feature which could
significantly inhibit the channel suggested.  For example, the
\GW recoil produced during a BH-BH merger likely ejects
the products of such a merger from any protocluster and prevents
BH-BH merger runaway; the characteristic \GW merger
time of evaporated BH binaries depends sensitively ($\propto \sigma^8$) on the
inital cluster 
velocity dispersion; and even the initial cluster's dynamics  differs
significantly depending on whether a discrete (two-component) or
realistic (continuous) mass distribution is used.

\begin{figure}
\includegraphics[width=\columnwidth]{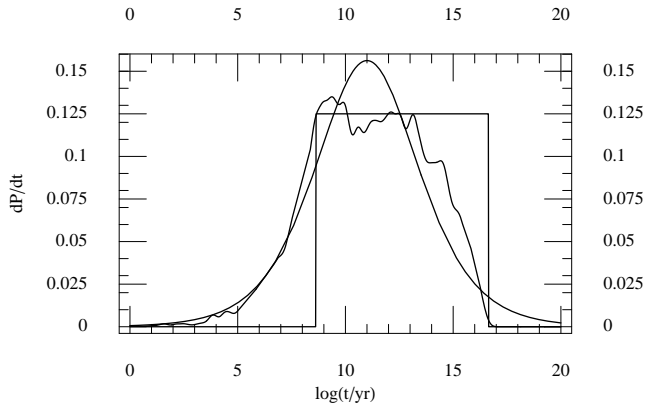}
\caption{\label{fig:mergerDelayTimes}
The distribution of \GW merger timescales for  circular
(solid step function)
and eccentric (solid smooth function, based on $10^4$ samples)  $14M_\odot+14 M_\odot$
binaries, assuming the binding energy  is distributed logarithmically
from $10^3 k_b T$ to $10^5 k_b T$ for $k_b T$ the characteristic
cluster energy.  This figure assumes $k_b T = 0.5 M_\odot (10 \unitkms)^2/3$.
}
\end{figure}

For this reason,  \citet{2006ApJ...637..937O} performed a broad range
of detailed 
numerical simulations of the dynamics of the segregated core and its
coupling to the surrounding cluster, incorporating  several critical physical
features of dense cluster dynamics (a realistic black hole
mass distribution; a range of BH-BH merger kicks that includes recoil
speeds larger than the most extreme cluster escape velocities, as
suggested by the most recent numerical simulations of unequal-mass binaries
\cite{2006.gr-qc..0610154,2006.gr-qc..0601026}; a range of cluster
velocity dipersions $\sigma$ from $5$ to $20 \unitkms$, encompassing the
range seen in present-day globular clusters; and both three-body and four-body
interactions). 
%% FRACTION OF BINARIES EJECTED
These simulations form the basis of our estimates of the numbers and
properties of the evaporated BH-BH binary population  
($N_{\rm bin} \approx 0.07 N_{\rm   bh}$)  that is ejected during the decoupled phase,
 consistent with theoretical expectations \citep{1993Natur.364..423S,1993Natur.364..421K}.
%% PROPERTIES of the hot binaries
These ``hot'' ejected binaries have a thermal eccentricity  ($e$)
distribution ($2 e de$)  and a lognormal  binding energy ($E_b = G m_1 m_2/2 a$)
distribution, with half the binaries having binding energies between
$10^{3.5}$ and $10^{4.5}$ times the thermal energy of the core, $kT =
\left< m_* \right> \sigma^2/3$ (see Fig. 6 of O'Leary et al.).    
Once ejected, the BH-BH binaries'
orbits decay through \GW emission, according to
Peters' Eqs. (5.6-7) \cite{Peters:1964}, until eventually each binary
merges.  The initially most strongly bound and eccentric orbits merge first,
followed by the initially widest and most circular
[Fig. \ref{fig:mergerDelayTimes}], with a \GW delay
time $t_{gw}\propto a^4$.  Furthermore, because of the  broad range of
ejected binary energies,  while the differential distribution's
\emph{characteristic} merger time depends sensitively on the assumed cluster velocity
dispersion ($t_{gw}\propto a^4 \propto \sigma^{-8}$) and to a lesser
extent on the precise details of the energy spectrum of ejected
binaries, the delay time 
distribution between $0.1$ and $13$ Gyr is comparatively robust,
varying by $\sim \, 50\%$.
Specifically, when a  population of
$m_{bh}=14 M_\odot$ black holes evaporates from inside a cluster with  a
$\sigma_{\rm cl}=10 \unitkms$ velocity dispersion, the binary black holes will have a
probability $P(<t)$ of merging by time $t$, with
\begin{equation}
\label{eq:rate}
\frac{dP}{dt}(t) \approx \frac{2.5}{16 t \log(10) } 
   {\rm sech} \left[ \frac{2.5(\log(t)-11)}{8}
   \right],
\end{equation}
where we simplify the distribution of binding energies by assuming
that it is flat in the log; a similar expression applies to general velocity dispersions
$\sigma_{\rm cl}$, when
the logarithm (only) is rescaled according to 
$\log t\rightarrow \log t(\sigma_{\rm cl}/10 \unitkms)^8$.  Except for very
early and very late times, this distribution is well approximated by
$dP/dt \approx  0.054/t$, as expected from the relation between
\GW delay time ($t_{gw}\propto a^4$) and the
logarithmic distribution of semimajor axes implied by cluster evaporation.
This simple   $1/t$ decay in merger rate per cluster appears in
the simulations of  O'Leary et al. (see their Figure 5).

\begin{figure}
\includegraphics[width=\columnwidth]{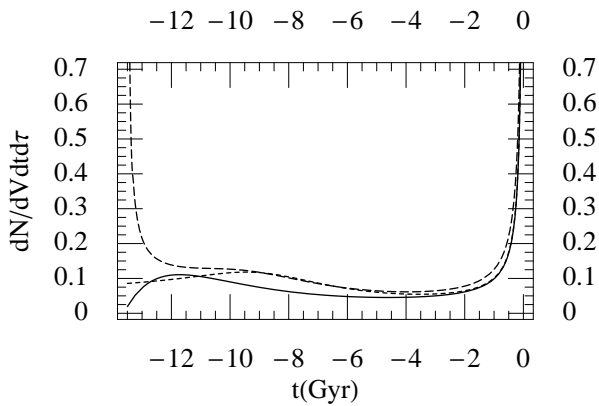}
\caption{\label{fig:earlyTimesMatterToo}
The distribution of binary birth times for those binary black holes
merging at present, given three different assumptions for the star
formation history in the universe: 
\citet{2006.astro-ph..0603257} (solid),
Porciani and Madau's 
Eq. (5)  (dotted),  % PM2
and (6) (dashed)    % PM3  (icky)
\cite{PorcianiMadau}.  In all three cases,
clusters formed more than $5$ Gyr ago produce more  BH-BH mergers than
more recent cluster formation.
}
\end{figure}

Combining the number of black holes we expect per unit mass ($N_{bh}
\approx 3\times 10^{-3} (M_{cl}/M_\odot)$), the fraction of those
black holes which should be ejected as binaries ($N_{\rm bin}\approx
0.07 N_{\rm bh}$, based on a variety of simulations \cite{2006ApJ...637..937O}),
and the rate of mergers per unit time given by Eq. (\ref{eq:rate}), we
expect the number of cluster mergers 
per unit cluster mass after a time $t$ to be ${\cal R}_{cl}(t) =
2.1\times 10^{-4} M_\odot^{-1}\, dP/dt $.  Assuming that clusters are
formed from a fraction 
$\gcl$ of available star formation and that only a fraction
$\gcand$ of all cluster-forming mass possesses the birth conditions 
necessary for this process to occur, then the BH-BH merger rate
$R_{\rm evap}(t)$  per
unit comoving volume  is
\begin{equation}
R_{\rm evap}(t) = \int^t d\tau \gcl \gcand 
 {\cal R}_{cl}(t-\tau)
 \frac{d\rho}{dt}(\tau) \, .
\end{equation}
where $d\rho/dt$ is the observed star formation rate per unit volume
in the universe  \cite{2006.astro-ph..0603257,2006.astro-ph..0611283}.
While observations have yet to converge to a unique star formation
history, most models for $d\rho/dt$ imply a present-day merger rate
within $30\%$ of $\approx \gcand \gcl {\rm Mpc}^{-3} {\rm
  Myr}^{-1}$; they also universally imply these binaries are born
equally often in the nearby and early universe
[Fig. \ref{fig:earlyTimesMatterToo}]. 
Finally, based on the  weak scaling of merger rate with
$\sigma_{\rm cl}$
implied by Peters' equations, the BH-BH merger rate 
depends weakly (empirically, roughly \emph{logarithmically}) on the assumed initial cluster velocity
dispersion 
$\sigma_{\rm cl}$ between $5{\rm km/s}$ and $20 {\rm km/s}$.  In
short, binary BHs merging due to this process do so at a relatively
well-defined rate, agreeing to order of magnitude with the merger rate
expected if all mass presently in stars  
formed  $10$ Gyr ago: $R_{\rm
  evap} \approx \gcl \gcand 2\times 10^{-5} M_{\odot}^{-1} \rho_*/(10
Gyr)\approx 0.8  \gcand \gcl {\rm Mpc}^{-3} {\rm
  Myr}^{-1}$.\footnote{The mass density in stars $\rho_*$ is roughly $0.3\%$ of the
closure density of the universe $\rho_c= 3H_o^2/8\pi G$
\cite{2004ApJ...616..643F}.}
%,2005RSPTA.363.2693R,WMAP-Year3Cosmology}.]  

%\section{Detection and comparison}
% POINT: Higher than popsyn --> constrainable
Unless suppressed strongly (i.e., $\gcl \gcand \ll 10^{-2} $), the BH-BH
merger rate due to clusters will significantly exceed the average rate densities expected from isolated stellar
evolution, \editremark{$R_{\rm popsyn}$} $\approx 10^{-2} {\rm Mpc}^{-3} {\rm
  Myr}^{-1}$  \cite{StarTrack,PSmoreconstraints}.  Additionally, as
noted in O'Leary et al., the BH-BH binaries produced from dynamical
cluster evaporation will strongly favor pairs of the highest-mass
binaries (e.g., $14M_\odot+14M_\odot$ or even $20M_\odot+20M_\odot$).
On the other hand, isolated binaries rely on \SN
kicks and mass transfer to bring them to merging; therefore these
merging BH-BH binaries are typically significantly less massive.
Because merging BH-BH binaries produced in  clusters possess a
distinctive mass signature and could occur at unusually high rates,
\GW observatories can directly constrain or even
measure $\gcl \gcand$.
For example, in an optimistic case ($\gcl=\gcand=1$), the initial
LIGO network  could detect roughly $10$ events per year, based on an
estimated network 
range to $14 M_\odot+14M_\odot$ binaries of $125 {\rm Mpc}$; the
``enhanced LIGO'' upgrade, with roughly twice the sensitivity, should
see roughly $2^{15/6}$ as many sources; and with 
roughly $20\times$ the range, the
advanced LIGO network could detect as many as  $3\times 10^4 \gcl
\gcand$ sources per year, permitting exquisite probes of
early-universe cluster dynamics if indeed this process yields more
BH-BH mergers than isolated stellar evolution.%
\footnote{The factor relating the initial and advanced LIGO detection rates is not
precisely geometrical (i.e., $20^{15/6}$) due  to cosmological
redshift of the emitted gravitational waves
out of LIGO's sensitive  band, as well as cosmological volume factors influencing the
scale of the light cone near $z\approx 0.5$
\cite{1998PhRvD..57.4535F}; see  Eqs. (13-20) of
O'Leary et al. for details \cite{2006ApJ...637..937O}.}
\optional{
These rates are conservative: BHs as large as $20 M_\odot$ should be
present.  (ACTUALLY A LITTLE TRICKY).  SHOULD include a bit on Ryan's
actual simulations...sigh
}

Gravitational wave observatories provide useful information precisely
because $\gcl$ and $\gcand$ are so weakly constrained electromagnetically.
For example, based on galactic \cite{1991ApJ...371..171L,2003ARAA..41...57L} 
and extragalactic
\cite{1998ASPC..148..127H,2005ApJ...619..779C,2005ApJ...631L.133F}
cluster observations,  the fraction of stars born in clusters $\gcl$ could be
anywhere from $20\%$ to $100\%$. 
% POINT: Mass fraction
Similarly, observations of galactic and extragalactic clusters
cannot rule out all clusters more massive than $3\times 10^4 M_\odot$
and thus with more than $100$ BHs undergoing runaway segregation and
evaporation.   While such clusters are exceedingly rare in number
($p(M)dM \propto M^{-2}$ \cite{1999ApJ...527L..81Z,2006astro.ph..9062L,2006ApJ...648..572E}), they likely contain a significant fraction
of all cluster-forming mass: assuming that clusters range in size from $30 M_\odot$
to $10^7 M_\odot$, roughly 45\% of all cluster-forming
mass lies in these most massive clusters.   While this fraction
depends very weakly (logarithmically) on the limiting masses assumed
for the cluster mass spectrum, a flatter mass function,
as suggested by some observations \cite{2006astro.ph.11586D,2006astro.ph..9062L}, would
imply a significantly higher fraction of mass in these most massive
clusters.  Therefore, if all sufficiently massive clusters undergo
segregation runaway, $\gcand$ could be close to unity.
\editremark{Previous version: as high as $0.45$ or even $1$.  Based on
  the preferred number mentioned previously from the mass spectrum.}
% Big clusters: 2006ApJ...641..763W 
On the other hand, these evaporated BH-BH binaries appear  only if
their clusters of origin survived long enough as bound objects for
gravitational segregation to occur.  Because gravitational segregation
should occur much more rapidly than the age of observed long-lived
clusters, such as the globular clusters of the Milky Way,  $\gcand$ must be larger
than the corresponding fraction of a galaxy's mass: $\gcand \gtrsim 10^{-4}-10^{-3}$  \cite{2005ApJ...631L.133F}.
On the other hand,  gravitational segregation should at best compete
with and more likely occur  more   slowly than  ``infant
mortality,'' the tendency of roughly $70-90\%$  of young clusters to
disrupt within their first $\sim 10 \, {\rm Myr}$ due
to photoionization- and \SN-driven gas ejection 
\cite{2003ARAA..41...57L,2005ApJ...631L.133F,2005AA...441..117L,2005AA...441..949G,2002MNRAS.336.1188K,2006MNRAS.369L...9B}.
Based on only $10\%$ of all clusters surviving a rapid ``infant
mortality'' epoch and on $45\%$ of all clusters being sufficiently
massive for the Spitzer instability to occur, we expect $\gcand
\approx 5\times 10^{-2}$. 
To summarize, we expect $\gcl \gcand$ could be as high as $1$
(corresponding to 10
merger detections  per year with initial LIGO); is likely $5\times 10^{-2}$ (1
event every two years with initial LIGO; 3 events per year with
``enhanced'' LIGO); and is very likely higher than $10^{-4}$
(producing merging BH-BH binaries slightly less frequently than isolated
binary stars, but with systematically higher masses).

\begin{acknowledgements}
This work was supported by NSF grant PHY-0601995.  RO'S acknowledges support on
NSF grant PHYS-0353111 and  AST-0449558.
\end{acknowledgements}

\nocite{2006ApJ...637..937O}  % Ryan's cluster paper
\nocite{WMAP-Year3Cosmology}  % Cosmology reference
\nocite{2006ApJ...652.1129F}  % Review article on cluster
                              % observations...Fall...talking about
                              % infant mortality, etc
\nocite{2006astro.ph..6036S}  % Another paper on cluster observations
                              % (may change)
\nocite{2006astro.ph..9669B}
\bibliography{paper}
% long-grbs has the PM paper

\end{document}